\documentclass[aps,superscriptaddress,showpacs,showkeys,nofootinbib]{revtex4}
\usepackage{graphicx}

\newcommand{\eg}{e.g.,\,}
\newcommand{\ie}{i.e.,\,}
\newcommand{\be}{\begin{equation}}
\newcommand{\ee}{\end{equation}}
\newcommand{\bea}{\begin{eqnarray}}
\newcommand{\eea}{\end{eqnarray}}
\newcommand{\etal}{et al.}

\newcommand{\E}{{\cal E}}
\newcommand{\D}{{\cal D}}

\newcommand{\lamzero}{{\lambdabar_0}}

\newcommand{\ra}{\rightarrow}

\begin{document}

\bibliographystyle{apsrev}

\title{Distance-redshift from an optical metric that includes absorption }

\author{B. Chen}
\email{Bin.Chen-1@ou.edu}
\author{R. Kantowski}
\email{kantowski@nhn.ou.edu}
\affiliation{Homer L.~Dodge Department~of  Physics and Astronomy, University of
Oklahoma, 440 West Brooks, Room~100,  Norman, OK 73019, USA}
\date{\today}

\begin{abstract}
We show that it is possible to equate the intensity reduction of a  light wave caused by weak absorption  
with a geometrical reduction in intensity caused by a ``transverse" conformal 
transformation of the spacetime metric in which the wave travels. 
We are consequently able to modify Gordon's optical metric to account for 
electromagnetic properties of ponderable material whose properties
include both refraction and absorption.  
Unlike refraction alone however, including absorption 
requires a modification of the optical metric that depends on the eikonal of the wave itself. 
We derive the distance-redshift relation from the modified optical metric for 
Friedman-Lema\^itre-Robertson-Walker spacetimes whose cosmic fluid has associated 
refraction and absorption coefficients. We then fit the current supernovae data and provide 
an alternate explanation (other than dark energy) of the apparent acceleration of the universe.

\end{abstract}



\maketitle

\section{Introduction}\label{sec:intro}

The concept of an optical metric was introduced  by Gordon \citep{Gordon} who proved that 
solutions to Maxwell's equations in a curved spacetime 
filled with a fluid whose electromagnetic properties are 
described by a real permittivity $\epsilon(x)$ and a real
permeability $ \mu(x)$ (and consequently by a real refraction index 
$n(x)=\sqrt{\epsilon\mu}$) could be found by solving a 
modified version of Maxwell's equations 
in a related optical spacetime with vacuum values 
for the permittivity and permeability, \ie with 
$\epsilon(x)=1$ and $\mu(x)=1$ [see Eq.\,(\ref{modified-Max})]. There is a
direct relation between solutions 
in these two spacetimes, the physical spacetime 
with an index of refraction and its physical metric, and the optical 
spacetime with $n=1$ and its optical-metric. 
Gordon's original metric accounted for refraction only. An open question 
was whether or not absorption could also be accounted for by modifying the 
spacetime geometry. In \citep{Kantowski2} we made such a modification  and 
showed that Gordon's optical metric could be generalized to include absorption 
by allowing the metric to become complex. In \citep{Kantowski2} we defined a 
complex refraction index $N=n+i\kappa,$ and distinguished two different cases, 
\ie strong and week absorption. In the case of strong absorption the real and imaginary parts of the 
index of refraction are of the same order ($\kappa\sim n$), 
the eikonal  $S(x^a)$ has an imaginary part, and the optical metric is complex. 
In the weak absorption case ($k\ll n$), the eikonal  $S(x^a)$ can be taken as real, and the real part of the 
optical metric (Gordon's original metric) remains as the significant geometrical structure. 
In this paper we show that for weak absorption another generalization of Gordon's 
metric exists that can account for absorption as well as refraction. 
The generalization amounts to a transverse conformal transformation 
(see Sec.\,\ref{sec:conformal}) of Gordon's original optical metric. 
This new optical metric remains real but does include absorption. 
Including absorption via a transverse conformal transformation 
requires an optical metric that depends on the eikonal of the wave itself.  

The contents of this paper are as follows: In the next section we define transverse conformal 
transformations.
In Sec.\,\ref{sec:GO} we consider the geometrical optics approximation 
and relate wave properties in the two spacetimes, 
the physical and optical. In Sec.\,\ref{sec:tau-sigma} we 
compare the effects of the transverse conformal transformation with absorption 
and relate the conformal factor $\sigma$ to optical depth $\tau.$ 
In Sec.\,\ref{sec:RW} we construct the optical metric that accounts for 
absorption in Friedman-Lema\^itre-Robertson-Walker (FLRW) spacetimes.
In Sec.\,\ref{sec:generalization} we include refraction by generalizing Gordon's 
metric \citep{Gordon} to including absorption. 
In Sec.\,\ref{FLRW} we derive the refraction/absorption corrected distance-redshift 
relation from the generalized Gordon's optical metric in FLRW spacetimes. 
In Sec.\,\ref{sec:numerical} we fit the current supernovae data with the Hubble curve of a cosmological 
model in which the cosmic fluid has both an associated refraction index $n$ 
and a constant opacity parameter $\alpha,$ and provide an alternate 
interpretation of the apparent acceleration of the cosmic expansion. 
In Sec.\,\ref{sec:discussion} we give our conclusions.

\section{The ``Transverse" Conformal Transformation}\label{sec:conformal}

A conformal transformation \citep{Eisenhart}  
is defined as a rescaling of the metric and is 
usually written in a form similar to
\be\label{conf-global}
\hat{g}_{ab}=e^{2\sigma}g_{ab}
,\ee 
where $\sigma(x^a)$ is an arbitrary scalar 
function defined on the spacetime manifold.
The vacuum Maxwell equations\footnote{Square bracket $[\cdot]$ or parenthesis $(\cdot)$ symbolize 
respectively complete anti-symmetrization or symmetrization of the enclosed indices, and 
$\nabla$ with a subscript is covariant differentiation.} 
\bea\label{Max}
F_{[ab,c]}=0,\cr
\nabla_bF^{ab}=0,
\eea  
are formally invariant under a conformal transformation, \ie
\bea
\hat{F}_{[ab,c]}=0,\cr
\hat{\nabla}_b\hat{F}^{ab}=0,
\eea 
provided that
the covariant electromagnetic field tensor $F_{ab}$ transforms as
\be
\label{Global-Trans}
\hat{F}_{ab}=F_{ab},
\label{covariantF}
\ee
and hence the contravariant field transforms as
\be
\hat{F}^{ab}=e^{-4\sigma}F^{ab}.
\label{contravariantF}
\ee
The above is a purely mathematical observation and 
simply says that if you have a solution to Maxwell's 
vacuum equations in one spacetime then you have a 
related solution to Maxwell's vacuum equations in 
any conformally related spacetime. Nothing is being 
said by this about the existence of any new physically relevant 
field.  One can and does make good use 
of such transformations to compactify spacetimes 
and to analyze fields  at $``\infty"$, see \eg \cite{Penrose}. 
However, by modifying these conformal transformations to be ``transverse" [see Eqs.\,(\ref{sub-conf}) and (\ref{sup-conf})]
we are able to construct new and useful solutions to Maxwell's equations.

Because the two metrics are defined on the same 
differentiable manifold we can make unique 
correspondences between events, world lines, and 
various fields in these two spacetimes. 
We call the 
original metric and manifold physical 
spacetime and the manifold with the conformally transformed metric 
the conformal spacetime.   For example a normalized 
fluid 4-velocity, $\hat u^a(x)$, defined in the 
conformal spacetime would be related to the 
corresponding physical fluid 4-velocity by 
$\hat u^a(x)= e^{-\sigma}u^a(x)$. If we 
compare observed properties of radiation fields that are
related as in Eqs.\,(\ref{covariantF}) and (\ref{contravariantF})
and as seen in these two spacetimes by corresponding 
observers, we find that  energy fluxes are diminished by a factor of $e^{-4\sigma}$. 
This net effect can be looked at as the result of the area of the radiation's 
wave front changing 
by a factor   $e^{2\sigma}$, and the energies and 
rates each changing by a factor  $e^{-\sigma}$. 
We will make use of this intensity reduction to account 
for absorption in physical spacetime. However, because a pure conformal
transformation has the undesirable property of altering a wave's frequency and wavelength
it must be appropriately modified to represent absorption, which obviously does not.

We assume that our physical spacetime is filled with two vector fields. The first is the normalized 
velocity field $u^a$ ($u^au_a=-1$) of a fluid whose linear and isotropic optical properties 
we know and the second is a null field $k^a$ ($k^ak_a=0$) 
corresponding to the eikonal of a given electromagnetic wave 
(see Sec.\,\ref{sec:GO} for details of the geometrical optics approximation).
These two fields define a 2-dimensional timelike subspace at each point of the manifold.  
We are hence able to decompose the tangent space at each point of the 4-dimensional spacetime manifold into two orthogonal 2-dimensional subspaces, 
one timelike $g_{\parallel ab}$ and spanned by the pair $(u^a, k^a)$ and the other spacelike $g_{\perp ab}$. 
The full metric can be decomposed in terms of these two orthogonal pieces as
\be
g_{ab}=g_{\parallel ab}+g_{\perp ab}.
\ee
We can give a simple expression for $g_{\parallel ab}$ by defining 
a second null vector $\ell^a$ lying in the two dimensional 
timelike subspace spanned by $u^a$ and $k^a$ as 
\be\label{def-m}
\ell^a\equiv {k^a\over 2(u\cdot k)^2}+{u^a\over (u\cdot k)},
\ee 
from which it follows that
\be
\ell^a\ell_a=0,\> \ell^au_a=-{1\over 2(u\cdot k)},\> \ell^ak_a=1.
\ee 
The metrics on the orthogonal 2-dimensional surfaces can be written as
\bea
g_{\parallel ab}&=&2\ell_{(a}k_{b)},\nonumber\\
g_{\perp ab}&=&g_{ab}-2\ell_{(a}k_{b)}.
\eea
We  conformally transform the $2$-dimensional spacelike subspace only
and arrive at the desired
 optical metric
\bea
\label{sub-conf}
\tilde{g}_{ab}&\equiv& g_{\parallel ab}+e^{2\sigma}g_{\perp ab},\nonumber\\
&=& e^{2\sigma}g_{ab}+(1-e^{2\sigma})\, 2 \ell_{(a}k_{b)},
\eea 
with inverse
\bea
\tilde{g}^{ab}&\equiv& g_{\parallel}^{ab}+e^{-2\sigma}g_{\perp}^{ab},\nonumber\\
&=&e^{-2\sigma}g^{ab}+(1-e^{-2\sigma}) 2 \ell^{(a}k^{b)}.
\label{sup-conf}
\eea
By straightforward tensor algebra we find that
\be
\det \tilde{g}=e^{4\sigma} \det g.
\ee 
We would have obtained $\det \hat{g}=e^{8\sigma}\det g$ were we conformally transforming 
the entire metric $g_{ab}$ as in Eq.\,(\ref{conf-global}).
We call these transformations transverse-conformal 
because they scale directions transverse to a wave's propagation 
direction $k^a$ as seen by an observer $u^a$ moving with the optical fluid.
 We find that the vacuum Maxwell equations (\ref{Max})  
remain invariant in form under the transverse conformal 
transformation of Eq. (\ref{sub-conf}), \ie
\bea
\tilde{F}_{[ab,c]}=0,\cr
\tilde{\nabla}_b\tilde{F}^{ab}=0,
\eea 
provided that
the covariant electromagnetic field tensor $F_{ab}$ transforms as
\be\label{conf-cond}
\tilde{F}_{ab}=F_{ab},
\ee
and that the contravariant field tensor $\tilde{F}^{ab}$ defined by
\be
\tilde{F}^{ab}\equiv \tilde{g}^{ac}\tilde{g}^{bd}\tilde{F}_{cd}
\label{tildeF}
\ee
is constrained to satisfy
\be
\tilde{F}^{ab}=e^{-2\sigma}F^{ab}.
\ee
 This constraint requires that two of the following three terms, \ie $F_0^{ab}$ and $F_2^{ab},$ originating 
from Eq.\,(\ref{tildeF}) vanish:
\be\label{raise-index}
\tilde{F}^{ab}=F^{ab}_0+e^{-2\sigma}F_1^{ab}+e^{-4\sigma}F_2^{ab},
\ee where
\bea
F_0^{ab}&\equiv& 2\ell^{[a}k^{b]}\left(F_{cd}k^c\ell^d\right),\cr
F_1^{ab}&\equiv& -2k^{[a}F^{b]}_{\ \ c}\ell^c-2\ell^{[a}F^{b]}_{\ \ c}k^c-2 F_0^{ab},\cr
F_2^{ab}&\equiv& F^{ab}-F_0^{ab}-F_1^{ab}.
\eea 
Consequently  $F_{ab}$ must satisfy 
\be\label{constraint-eq-a}
F_{cd}k^c\ell^d=0,
\ee and
\be\label{constraint-eq-b}
F^{ab}=-2k^{[a}F^{b]}_{\ \ c}\ell^c-2\ell^{[a}F^{b]}_{\ \ c}k^c.
\ee  

As we will see  in the next section [see Eq.\,(\ref{F})] a radiation field whose eikonal 
generates the null field $k^a$ satisfies these constraints. 
For a familiar example, choose the physical metric to be Minkowskian, \ie
\be
ds^2=-c^2dt^2 +dr^2+r^2(d\theta^2+\sin^2\theta\, d\phi^2), 
\ee 
with a stationary optical fluid  $u^a=\left(1/c,0,0,0\right)$, and radial null geodesics $k^a=k\left(1/c,1,0,0\right)$. From Eq.\,(\ref{def-m}) we find
\be
\ell^a={1\over 2k}\left(-{1\over c},1,0,0\right).
\ee 
Equations\,(\ref{constraint-eq-a}) and (\ref{constraint-eq-b}) then give 
\be\label{Trans-Cond}
F_{01}=-F_{10}=0, \> F_{23}=-F_{32}=0,
\ee 
\ie $F^{ab}$ is a transverse field, hence motivating the designation of  
Eq.\,(\ref{sub-conf}) as a ``transverse" conformal transformation.
 Equations (\ref{constraint-eq-a}) and (\ref{constraint-eq-b}) simply express 
the transversality condition for an arbitrary wave. 
In the following section we review the geometrical optics approximation 
and relate electromagnetic quantities in the 
physical and optical spacetimes.

\section{Geometrical Optics Approximation} \label{sec:GO}

Following \citep{Ehlers2, Sachs0, Kantowski2} we write the covariant (and metric independent) 
field tensor as\be\label{WKB-1}
F_{ab}={\rm Re}\left\{e^{iS/\lamzero}\left(A_{ab}+{\lamzero\over i}B_{ab}+O(\lamzero^2)\right)\right\}, 
\ee 
where $\lamzero$ is a wavelength related parameter, $S(x^a)$ is the so-called eikonal function and is real,  
and ${\rm Re}\{\cdot\}$ stands for the real part. The $A_{ab}$ term represents the geometrical 
optics approximation and the $B_{ab}$ term is its first order correction.
Defining the unitless (also metric independent) wave vector $k_a=\partial_a S$ 
and inserting Eq.\,(\ref{WKB-1}) 
into the vacuum Maxwell equations we obtain to order $\lamzero^{-1}$ 
\bea\label{0th-a}
A_{[ab}k_{c]}&=&0,\cr
A^{ab}k_b&=&0,
\eea 
and to order $\lamzero^0$
\bea\label{1st-a}
\partial_{[a}A_{bc]}+k_{[a}B_{bc]}&=&0,\cr
\nabla_bA^{ab}+B^{ab}k_b&=&0.
\eea
Equations (\ref{0th-a}) tell us that $k^a\equiv g^{ab}k_b$ is tangent to 
null geodesics of $g_{ab}$,  \ie
\bea
k^ak_a&=&0,\nonumber\\
k^b\nabla_bk^a&=&0,
\label{Null}
\eea  and that $A_{ab}$ is of the form:
\be
A_{ab}=-2k_{[a}\E_{b]},
\label{GO-A}
\ee  
where $\E_a$ is spacelike and constrained by $\E_ak^a=0$ 
with the remaining gauge freedom (to order $\lamzero^0$) $\E_a\ra \E_a+f(x)k_a$.  
Since $k^a\ell_a=1,$  we can use this freedom to choose $\E_a$ such that $\E_a\ell^a=0$ 
or equivalently $\E_au^a=0$.
For this choice, $\E^a$ is (up to a factor $\omega^{-1}$) the amplitude of the electric field seen by observers at rest with respect to the fluid $u^a$. 
It is the geometrical optics approximation, \ie Eq.\,(\ref{GO-A}) 
that makes the Maxwell field of  Eq.\,(\ref{WKB-1}) satisfy the 
needed transverse constraints of Eqs.\,(\ref{constraint-eq-a}) and (\ref{constraint-eq-b}) 
and hence allows us to introduce the transverse conformal transformation of Eq.\,(\ref{sub-conf}). 
Before doing so we first finish the geometrical optics approximation for $F^{ab}$ in the physical spacetime.

Equation (\ref{1st-a}) tells us that the order $\lamzero^1$ correction to geometrical optics is of the form 
\be
B_{ab}=2(\E_{[a,b]}-k_{[a}\D_{b]}),
\ee
with a remaining gauge freedom, to $O(\lamzero^1$), $\D_a\ra \D_a+g(x)k_a$  
and also gives the propagation equation for $\E^a$ 
\be 
\dot{\E}^a+\theta\E^a={k^a\over 2}(\nabla_b\E^b+k_b\D^b),
\label{Edot}
\ee 
where `$\cdot$' is the affine parameter rate of change along the null ray,
$\dot{\E}^a\equiv k^b\nabla_b\E^a$,
and $\theta$ is the expansion rate of the null congruence, $\theta\equiv\nabla_ak^a/2$.
 By splitting $\E^a=\E e^a$ into a scalar amplitude $\E$ and a unit polarization 
vector $e^a\overstar{e}_a=1$ (where $*$ is complex conjugation), 
the transport equation for the amplitude $\E$  becomes 
\be
\dot{\E}+\E\theta=0.
\label{transport}
\ee 

 The geometrical optics approximation, \ie the $O(\lamzero^0)$ term in Eq. (\ref{WKB-1}), 
hence simplifies to
\be
F^{ab}=-2\,{\rm Re}\{\E\,e^{iS/\lamzero}k^{[a} {e}^{b]}\}.
\label{F} 
\ee

As we have indicated above, because both the physical metric $g_{ab}$ 
and optical metric $\tilde g_{ab}$ of Eq.\,(\ref{sub-conf}) are defined on the same 
manifold we can compare properties of a common field such as $F_{ab}$ in both spacetimes.  
For the above geometrical optics field all covariant quantities such 
as $\tilde F_{ab}, \tilde k_a, \tilde \ell_a,$ and $ \tilde\E_a$ in the 
optical spacetime are exactly 
the same as $F_{ab}, k_a,\ell_a,$ and $\E_a$ in the physical spacetime. 
All contravariant components that lie in the $k$-$\ell$  plane are also unchanged,
\ie  $\tilde k^a=k^a, \tilde \ell^a=\ell^a,$ and $\tilde u^a=u^a$. Consequently, quantities 
such as affine parameters, frequencies and wavelengths of the waves are the same in both spacetimes. 
However, transverse contravariant components,  \ie components in the orthogonal 2-dimensional
spacelike surface are scaled by the conformal factor 
$e^{-2\sigma}$, \eg $\tilde \E^a=e^{-2\sigma}\E^a$. 
This changes the magnitude to $\tilde \E=e^{-\sigma}\E$ with a new unit polarization 
vector $\tilde e^a=e^{-\sigma}e^a$
and the expansion parameter $\theta$ of Eq.\,(\ref{transport}) to $\tilde\theta=\theta+\dot\sigma$.

The time averaged 4-flux seen by an observer moving with the optical fluid is in general
\be
S^a\equiv \frac{c}{8\pi}{\rm Re}\left\{\overstar{H}^{ac}F_{cb}-\frac{1}{4}\delta^a_{\>b} \overstar{H}^{dc}F_{cd}\right\}u^b.
\label{4P}
\ee 
In the physical and optical spaces they are related by 
\be
\tilde{S}^a=e^{-2\sigma}S^a=-e^{-2\sigma}{c\over 8\pi}(u^bk_b)\E\overstar{\E}k^a,
\ee 
and from the $3$-D Poynting vector $S_{\perp}^a\equiv(g^{ab}+u^au^b)S_b$ we find magnitudes related by
\be\label{S-perp}
\tilde{S}_{\perp}=e^{-2\sigma}S_{\perp}=e^{-2\sigma}{c\over 8\pi}\E\overstar{\E}(u^bk_b)^2.
\ee
This result says that the intensity of a monochromatic wave can be reduced by a factor 
$e^{-2\sigma}$ at any point in spacetime by a transverse conformal transformation without 
altering the wave's frequency or wavelength. In the next section we equate 
this reduction of energy flux with absorption. 
 In Sec.~\ref{sec:generalization} when we combine absorption with refraction 
we reverse the process and assume that physical spacetime possesses the absorbing material and the transverse 
conformal transformation  to the optical spacetime removes it.

\section{A Transverse Conformal Transformation Versus Absorption}\label{sec:tau-sigma}

The attenuation coefficient $\kappa_\nu$ (${\rm cm}^2\cdot{\rm g}^{-1}$) is defined by looking at the amount of 
energy absorbed, $dE_\nu$, in time  $dt$  by a pencil beam of radiation 
as it passes through a small slab of material of density $\rho$ (${\rm g}\cdot{\rm cm}^{-3}$), 
cross-section $dA$ and length $dl$ (see e.g. \citep{Mihalas}) 
\be
dE_\nu=(\rho \kappa_\nu)I_\nu d\Omega d\nu dt dA dl.
\ee 
All quantities are defined in the  local co-moving frame of the optical fluid $u^a$.
The specific intensity $I_\nu$ ($\times d\Omega d\nu$) measures the wave's intensity 
directed into solid angle 
$d\Omega$ and within the frequency range $\nu$ to $\nu+d\nu.$   
In the absence of emission the specific intensity $I_\nu(\tau)$, after traveling 
an optical depth $\tau\equiv \int{\rho \kappa_\nu dl}$ along a 
``characteristic" direction, differs from the value $I_\nu(0)$ it would have  without absorption by
\be 
I_\nu(\tau)=I_\nu(0)e^{-\tau}.
\label{extinction}
\ee 
Evaluating  $\tau$ can be complicated because the frequency in the integrand is 
continually Doppler shifted due to the 
non-static nature of the optical fluid and/or the curvature of spacetime.
The specific flux $S_\nu$ at any point is the first moment of 
specific intensity $I_\nu$, \ie
\be
S_\nu=2\pi\int_{-1}^{1}{\mu I_{\nu} d\mu}.
\ee 
For any $I_\nu$ that has a delta function dependence on direction, 
\eg  a geometrical optics wave, 
 the corresponding value of the flux
$S_\nu(\tau)$ seen by the observer in the presence of absorption is similarly related 
to the non-absorption value, \ie 
\be
\label{S-tau}
S_\nu(\tau)=S_\nu(0) e^{-\tau}.
\ee 
Comparing Eqs.\,(\ref{S-perp}) and ({\ref{S-tau}}) we find that a conformal 
factor $\sigma$ reduces the 
flux by the exact same amount as absorption if
\be\label{sig-tau}
\sigma={1\over 2}\tau.
\ee 
If this conformal factor is to be unique, then either a single frequency is 
present at any spacetime point or 
the attenuation is ``grey", \ie the opacity $\alpha\equiv\rho\kappa_{\nu}$  
is frequency independent.  
Even if there is only a single frequency present the 
frequency dependence in the absorption makes the calculation 
complicated. We must follow the frequency of each wave, 
starting from the source,
as it is red/blue shifted and selectively absorbed while moving through the optical medium.
In general a different conformal factor would exist for each source frequency.
However, for the ``grey" case  the $\sigma(x^a)$  is unique.  
For a geometrical optics wave emanating from the world line of a point source, $\sigma(x^a)$ 
is defined on  forward light cones (surfaces on which the eikonal $S$ remains constant), \ie 
\be
\sigma(x^a)={1\over 2}\int_{x_s^a}^{x^a}\rho(x'^b) \kappa_\nu(x'^b)dl.
\ee 
The integration is performed along the null geodesic connecting the 
emitting event $x_s^a$ and the spacetime point $x^a$. 
The density, 
attenuation coefficient, and spatial length element, respectively 
$\rho, \kappa_\nu,$ and $dl$, are measured in a sequence 
of local inertial frames ($u^a\propto\delta^a_0$) which are at rest with respect to 
the material along the null geodesic. 
We caution the reader that $\sigma(x^a)$ might not be globally defined, 
or even uniquely defined. The integral above is only defined for $x^a$ 
which lie on  forward light cones of the source. 
If the light source is turned on/off at some time (star birth/death) 
$\sigma$ will only be defined on part of the spacetime manifold. 
However, its value can be taken as zero on the remainder.
Furthermore, we might have multiple null geodesics connecting the emitting 
event $S$ and receiving event $O.$ For an example, when an Einstein ring 
occurs in a gravitational lensing configuration we have, in principle, 
an infinite number of null geodesics connecting $S$ and $O.$ 
These rays might pass through 
regions with different $\rho$'s and $\kappa_\nu$'s and therefore give us different $\sigma$'s. 
Rather than limiting the domain over which the conformal transformation is defined to keep 
it single valued, we can extend the manifold  
to multiple layers and make the conformal factor unique  on each layer. 
This is in direct analogy with Riemann's extension of the domain of a 
multi-valued complex function on R$^2$ to a Riemann surface on which 
its value is unique. This extended domain could be truly convoluted as 
for example in an inhomogeneous cosmology where strong lensing is prevalent. 

The simplest example to illustrate a transverse conformal transformation 
is a plane electromagnetic 
wave traveling in Minkowski spacetime filled with stratified weakly absorbing 
gas. We suppose a monochromatic light source is lying infinitely far away ($z=-\infty$) 
and is producing a plane wave propagating in the $+z$ direction. 
The needed properties of a stratified gas filling the  spacetime are
summarized by the density $\rho(z)$ and the attenuation coefficient $\kappa_\nu(z).$ 
The physical metric is
\be
ds^2=-c^2dt^2+dx^2+dy^2+dz^2,
\label{flat}
\ee
and the vector fields $k^a$ and $\ell^a$ in Eq.\,(\ref{def-m}) are trivially ($u^a=\delta^a_t/c$) 
\bea
k^a&=&k\left({1\over c}, 0,0,1\right),\nonumber\\  
\ell^a&=&{1\over 2k}\left(-{1\over c}, 0,0,1\right).
\label{rays}
\eea
Equation\,(\ref{sub-conf}) then gives
\be
d\tilde{s}^2=-c^2dt^2+e^{2\sigma(z)}(dx^2+dy^2)+dz^2,
\label{conf-transed}
\ee 
where the conformal factor is related to the integral of the opacity by
\be
\sigma(z)={1\over 2}\int_{-\infty}^{z}\alpha(z')dz'.
\ee 
We can now give a geometrical  interpretation of Eq.\,(\ref{S-perp}). 
Because the rays in Eq.\,(\ref{rays}) are all running parallel to the z-axis 
the cross-sectional area of any bundle of rays is 
expanded by a factor $e^{2\sigma(z)}$ for the conformally transformed 
spacetime, Eq.\,(\ref{conf-transed}), relative to initial the physical spacetime, 
Eq.\,(\ref{flat}).
From the definition of flux, we can conclude that it will be reduced by 
the reciprocal factor $e^{-2\sigma}.$ The reason the conformal factor was not applied 
to the whole $4$-D metric now becomes clear: a $e^{2\sigma}$ 
factor in front of the entire physical metric would produce an undesirable time 
dilation factor $e^{\sigma}$ as well as a frequency shift factor $e^{-\sigma}.$ 
These two combined would have 
contributed another factor of $e^{-2\sigma}$ to the wave's flux in the conformal spacetime.
We can also look at the electric field ${\bf E}$ and  the magnetic field ${\bf H}$
in both spacetimes and see that 
in the conformal spacetime they are each diminished by a factor $e^{-\sigma}$.
Consequently the 
flux, ${\bf E}\times {\bf H}$, is reduced by $e^{-2\sigma}$. 
(This reduction in amplitude/intensity matches that of weak absorption given in \cite{Kantowski2} but, 
not unexpectedly, differs from that of strong absorption, \ie Eqs.~(28)-(30) of \cite{Kantowski2}).

\section{Absorption in FLRW  Spacetimes}\label{sec:RW}

The  interesting example for cosmology is absorption by the Intergalactic Medium (IGM) 
which is naturally modeled as absorption by the cosmological fluid in 
Friedman-Lema\^itre-Robertson-Walker (FLRW) spactimes.
 The familiar Robertson-Walker metric can be written in co-moving coordinates ($u^a=\delta^a_t/c$) as
\be
ds^2=-c^2\,dt^2+R^2(t)\left\{{dr^2\over 1-kr^2}+r^2(d\theta^2+\sin^2\theta\, d\phi^2)\right\},
\label{FRW}
\ee 
where $k=1,0,-1$ for a closed, flat or open universe, respectively. 
The eikonal of Eq.\,(\ref{F}) for a point source  located at $r=0$ (see \cite{Kantowski2}) is 

\be
S(t,r)=R_0\left(-\int^{t}_{t_e}{c\,dt'\over R(t')}+
{\rm sinn}^{-1}[r]\,\right),
\ee
where
\be
{\rm sinn}[r] \equiv \cases{\sin[r] & $k=+1$\cr r & $k=0$\cr \sinh[r] & $k=-1$,}
\ee 
and $R_0$  is the radius of the universe at the observation time $t_0$.
The corresponding radial null geodesics are
\be
k^a=R_0\left({1\over c\,R(t)},{\sqrt{1-kr^2}\over R^2(t)},0,0\right),
\label{null-geodesic}
\ee 
for which 
Eq.\,(\ref{def-m}) gives 
\be
\ell^a={R(t)\over 2R_0}\left(-{1\over c}, {\sqrt{1-kr^2}\over R(t)},0,0\right),
\ee 
and from which it follows that
\be
2\ell_{(a}k_{b)}=\hbox{diag}\,\left[-c^2, {R^2(t)\over 1-kr^2},0,0\right].
\ee Equation\,(\ref{sub-conf}) then gives the optical metric
\be\label{RW-Absorb}
d\tilde{s}^2=-c^2\,dt^2+R^2(t)\left\{{dr^2\over 1-kr^2}+e^{2\sigma}r^2(d\theta^2+\sin^2\theta\, d\phi^2)\right\}.
\ee 

An observer at $(r,t)$  will see the source  redshifted an amount $z$ 
as a function of co-moving radius $r$ and observing instant $t$, which can be obtained by integrating 
Eq.\,(\ref{null-geodesic}). If the dynamics of a Robertson-Walker metric 
is determined by general relativity, 
and the gravity source is a mixture of non-interacting perfect fluids 
including a cosmological constant, 
the cosmology is called FLRW and has a redshift which can be found by inverting
\be
r(t,z)={\rm sinn}\,\left[{c\over R(t)H(t)}\int_{0}^{z}{dz'\over h(z') }\right],
\ee where 
\be
h(z)=\sqrt{\Omega_\Lambda+\Omega_k(1+z)^2+\Omega_{\rm m}(1+z)^3+\Omega_r(1+z)^4}.
\ee 
Here $\Omega_\Lambda,\Omega_{\rm m},\Omega_r$ are the respective density parameters for the assumed vacuum energy, 
cold matter and radiation content  of the current universe.

For spatially homogeneous and nondispersive (grey) absorption the differential optical depth $d\tau$ is related to the opacity $\alpha$ by
\bea
d\tau&=&-\alpha (cdt)\cr
&=& \alpha {c\over H}{1\over (1+z)}{dz\over h(z)},
\eea 
and the total optical depth $\tau(r,t)$ of a source at $(0,t_e(r,t))$ 
seen by an observer at $(r,t)$ is the integral
\be\label{RW-tau}
\tau(z)={c\over H(t)}\int_{0}^{z}{{\alpha(z')\over (1+z')h(z')}dz'}.
\ee 
The conformal factor $\sigma$ of Eq.\,(\ref{RW-Absorb}) is therefore one-half this value.
 
The presence of absorption changes the distance modulus-redshift relation $\mu(z),$ e.g., 
the magnitudes of supernovae are corrected for absorption by the IGM before 
drawing a Hubble diagram.  From the definition of distance modulus
\be
\mu\equiv5\log {d_L\over 1 \hbox{Mpc}}+25,
\ee 
and luminosity distance $d_L$ in terms of flux $S$ 
\be
d_L=\left({L\over 4\pi S}\right)^{1/2},
\ee 
we immediately obtain via Eq.\,(\ref{S-tau})  the absorption-corrected luminosity distance
\be\label{DL-Absorb}
\tilde{d}_L=e^{\tau/2} d_L,
\ee 
and the corrected distance modulus $\tilde{\mu}$
\be
\tilde{\mu}=\mu+{5\over 2\ln 10} \tau.
\label{mu-tau}
\ee 
This is the normal interpretation of the increase in luminosity distance 
caused by light absorption. If we use the optical metric to compute luminosity distance (see e.g. \citep{Ellis}) we find that compared 
with the physical metric the cross-sectional area of the ray bundle is expanded by a factor 
$e^{2\sigma}$  which reduces the received flux by a factor of $e^{-2\sigma}$, \ie  
from Eq.\,(\ref{RW-Absorb})  
\be
\tilde{d}_L=e^{\sigma(r_o,t_o)}(1+z)R_0r=e^{\sigma}d_L.
\ee We see that by identifying $\sigma$ with $\tau/2$ 
we have the same luminosity distance in the optical spacetime with no absorption 
as in the physical spacetime with absorption.

\section{Incorporating Both Refraction and Absorption into the Optical Metric }\label{sec:generalization}

Up to now, we have been considering only absorption, however,  
we know that besides having its flux reduced, a light wave's path and speed 
can be altered (refracted) due to the presence of a polarizable material. 
An interesting and useful tool to study refraction in curved spacetime was 
developed by Gordon \citep{Gordon} (see also Ehlers \citep{Ehlers1}). 
Gordon modified Einstein's physical spacetime metric to include the effects 
of a nondispersive refractive material on Maxwell's theory.   
The theory accounts for any polarizable  material whose 
constitutive properties are summarized by two scalar functions, 
a permittivity $\epsilon(x^a)$ and a permeability $\mu(x^a). $
His optical  metric $\bar{g}_{ab}$ \cite{Gordon} 
is defined on the same differentiable manifold as Einstein's spacetime metric $g_{ab}$ 
and is related to it by
\be
\bar{g}_{ab}=g_{ab}+\left(1-{1\over n^2}\right)u_au_b,\>\>\> \bar{g}^{ab}=g^{ab}+\left(1- n^2\right)u^au^b,
\label{Gordon}
\ee 
where $n(x^a)=\sqrt{\epsilon\mu}$ is the refractive index and 
$u^a$ is the 4-velocity of the optical fluid, 
normalized using the physical spacetime metric $g_{ab}$.
Null geodesics of the $\bar{g}_{ab}$ metric are identical to
light paths (timelike) in physical spacetime filled with a refracting material of index $n(x)$. 
Even though we can incorporate absorption of only geometrical optics waves
into Gordon's metric, his index of refraction theory applies to all Maxwell fields, 
see \cite{Kantowski2} for some related examples.  
Gordon showed that solutions to Maxwell's theory ($F_{ab},H^{ab}$) in the presence of such a material 
can be found by solving a slightly modified set of vacuum ($\epsilon=\mu=1$) 
Maxwell equations for ($\bar{F}_{ab}, \bar{F}^{ab}$)
 in his optical spacetime. The fields are connected by 
\bea
 H^{ab}&=&\frac{1}{\mu}\bar{F}^{ab},\nonumber\\
F_{ab}&=&\bar{F}_{ab},
\label{H}
\eea
and the modified Maxwell equations are \footnote{We thank John Ralston for pointing out to us that the modified equations can be derived from the Lagrangian ${\cal L}=-\frac{1}{4}\sqrt{-\det\bar{g}}\sqrt{\epsilon/\mu}\bar{F}^{ab}F_{ab}.$} 
\bea
\bar{F}_{[ab,c]}=0,\nonumber\\
\bar{\nabla}_b\left(\sqrt{\epsilon/\mu}\,\bar{F}^{ab}\right)=0.
\label{modified-Max}
\eea
We now generalize Gordon's optical metric to include effects of 
an additional isotropic and frequency independent (grey) attenuation coefficient $\kappa_\nu$ 
on a radiation field described by the geometrical optics approximation. To obtain this 
result we simply apply a transverse conformal transformation  
of Eq.\,(\ref{sub-conf}) to Gordon's metric Eq.\,(\ref{Gordon}).
The geometrical optics theory in  Gordon's optical spacetime is almost identical to the 
geometrical optics theory of the physical spacetime 
given in Sec.\,\ref{sec:GO} (see Sec.\,III of \cite{Kantowski2}), \eg
the contravariant components of the Maxwell field are correctly given by Eq.\,(\ref{F}) 
\be
\bar{F}^{ab}=-2\,{\rm Re}\{\E\,e^{iS/\lamzero}\bar{k}^{[a} \bar{e}^{b]}\}.
\label{Fbar}
\ee
Here  
\be
\bar{k}^a\equiv\bar{g}^{ab}\partial_bS,
\ee
 is tangent to null geodesics 
of the optical metric (and tangent to the corresponding timelike ``slower than light" curves of the physical metric), 
$\bar{e}^b$ is the unit (in the optical metric) polarization vector, and $S$ is the eikonal function. 
The significantly different equation is the transport equation for the amplitude of the wave given by 
Eq.\,(\ref{transport}), which becomes 
\be
\dot{\E}+\theta\E+\dot{\phi}\E=0,
\label{transport-bar}
\ee 
(see Eq.\,(18) of \cite{Kantowski2}). The presence of the additional term 
$\phi\equiv (1/4)\ln (\epsilon/\mu)$ 
is due to the afore mentioned modification of Maxwell's equations 
[see Eq.\,(\ref{modified-Max})]
in Gordon's optical spacetime.

From Eqs.\,(\ref{sub-conf}) and (\ref{sup-conf}) the  new optical metric becomes
\bea\label{GO-Sig}
\tilde{g}_{ab}=e^{2\sigma}\bar{g}_{ab}+(1-e^{2\sigma}) 2\bar{\ell}_{(a}\bar{k}_{b)},\>\>\> \tilde{g}^{ab}=e^{-2\sigma}\bar{g}^{ab}+(1-e^{-2\sigma}) 2\bar{\ell}^{(a}\bar{k}^{b)}.
\eea 
where $\bar{\ell}^a$ from Eq.\,(\ref{def-m}) is a null vector field in  Gordon's optical spacetime defined by
\be\label{m-bar} 
\bar{\ell}^a={\bar{k}^a\over 2(\bar{u}_b\bar{k}^b)^2}+{\bar{u}^a\over (\bar{u}_b\bar{k}^b)},
\ee 
and $\bar{u}^a$ the fluid's $4$-velocity normalized using  Gordon's optical metric
\be
\bar{u}^a=n u^a,\>\>\bar{u}_a={1\over n} u_a.
\ee 
The reader should observe that the transverse conformal transformation does not alter the 
orthogonality of the two 2-D subspaces  nor does it alter the metric structure 
of the timelike 2-D space spanned by $\bar{u}^a$ and $\bar{k}^a$. 
Rewriting the  new optical metric Eq.\,(\ref{GO-Sig}) in terms of the physical metric $g_{ab}$ 
and physical observer $u_a$ we have
\be
\tilde{g}_{ab}=e^{2\sigma} g_{ab}+e^{2\sigma}\left(1-{1\over n^2}\right)u_au_b+{1-e^{2\sigma}\over n^2(u\cdot k)^2}
\left[k_ak_b+2(u\cdot k)k_{(a}u_{b)}\right],
\ee 
with inverse
\be
\tilde{g}^{ab}=e^{-2\sigma}g^{ab}+(1-n^2)e^{-2\sigma}u^au^b+{1-e^{-2\sigma}\over n^2(u\cdot k)^2}
[\tilde{k}^a\tilde{k}^b+2n^2(u\cdot k)u^{(a}\tilde{k}^{b)}],
\ee 
where 
\be 
\tilde{k}^a\equiv \tilde{g}^{ab}S_{,b}=\bar{g}^{ab}S_{,b}=g^{ab}S_{,b}+(1-n^2)u^a(u^bS_{,b}),
\label{tildek}
\ee
and determinant
\be
\det \tilde{g}={e^{4\sigma}\over n^2}\det g.
\ee 
The new Maxwell field 
\bea
\tilde{F}^{ab}&=&\mu e^{-2\sigma}H^{ab},\cr
\tilde{F}_{ab}&=&F_{ab},
\eea 
satisfies the same equations as $\bar{F}^{ab}$, \ie\,Eq.\,(\ref{modified-Max}),
but $\tilde{g}_{ab}$ has the advantage of incorporating both refraction and absorption.
We now have a correspondence of geometrical optics waves in two spacetimes, the physical and the optical. 
In the physical spacetime the wave travels at a reduced speed $c/n$ with an intensity 
that is reduced by absorption as in Eq.\,(\ref{extinction}) whereas in the optical spacetime the wave
travels at speed $c$ with no extinction.

\section{The Optical Metric in FLRW Cosmology with Both Refraction and Absorption}\label{FLRW}

The formalism developed above appears complicated 
when applied to an arbitrary spacetime, however, for specific cases the formalism is more transparent.  
For example we reconsider FLRW spacetimes, but this 
time with both refraction $n$ and absorption $\kappa$ associated with the cosmic fluid. 
The result is elegantly simple. 
Gordon's optical metric for the Robertson-Walker metric \citep{Kantowski2} is 
\be
d\bar{s}^2=-{c^2\over n^2}dt^2+R^2(t)\left\{{dr^2\over 1-kr^2}+r^2(d\theta^2+\sin^2\theta\, d\phi^2)\right\}.
\label{OM-FRW}
\ee 
This metric does not measure distance and time, that is reserved for Eq.\,(\ref{FRW}). 
A dynamical theory that produced a gravitational field containing the 
index of refraction as given in Eq.\,(\ref{OM-FRW}) would be 
quite strange in that it would be partially sourced by electromagnetic polarization densities; 
which is not stress, energy, or momentum. The significant property of the optical metric 
that we use here is that its null geodesics are identical with the speed c/n photons 
of Eq.\,(\ref{FRW}).

For spherical waves the radial null geodesics are
\be
\bar{k}^a=R_0\left({n\over c\,R},{\sqrt{1-kr^2}\over R^2},0,0\right),
\label{kbarup}
\ee  and from Eq.\,(\ref{m-bar}) we find
\be
\bar{\ell}^a={R\over 2R_0}\left(-{n\over c},{\sqrt{1-kr^2}\over R},0,0\right).
\ee 
Immediately we find
\be
2\bar{\ell}_{(a}\bar{k}_{b)}=\hbox{diag}\>\left(-{c^2\over n^2},{R^2\over 1-kr^2},0,0\right),
\ee which gives us the new optical metric
\be\label{RW-opt}
d\tilde{s}^2=-{c^2\over n^2}dt^2+R^2(t)\left\{{dr^2\over 1-kr^2}+e^{2\sigma}r^2(d\theta^2+\sin^2\theta\, d\phi^2)\right\}.
\ee 
With dynamics supplied by the FLRW solutions the scalar function $\sigma$ in the above is
\bea
\sigma(r, t)&\equiv& {1\over 2}\int^{t}_{t_e}{\alpha\ {c\,dt'\over n}}\cr
&=&{1\over 2} {c\over H(t)}\int_{0}^{z(z_n)}{{\alpha(z')\over n(z')(1+z')h(z')}dz'}.
\eea Note the difference between wavelength redshift $z$ and frequency redshift $z_n$ (see \cite{Kantowski2}):
\be
1+z={R_0\over R(t_e)},\> 1+z_n={n(t_o)\over n(t_e)}{R_0\over R(t_e)}.
\ee 

We have discussed the impact of light refraction on the distance-redshift 
relation in \citep{Kantowski2}. Now incorporating absorption
we find from Eq.\,(\ref{RW-opt}) that the apparent size distance 
$\tilde{d}_A$ to the source in the new optical metric is
\be
\tilde{d}_A=R(t_e)e^{\sigma(0,t_e)}r=R(t_e)r=\bar{d}_A,
\ee 
is the same as the apparent size distance in Gordon's spacetime
\be\label{dA-z}
\bar{d}_A(z_n)={1\over [1+z(z_n)]}{c\over H_0}{1\over \sqrt{|\Omega_k|}}\hbox{sinn}
\left[ \sqrt{|\Omega_k|}\int^{z(z_n)}_0\frac{dz}{n(z)h(z)}\right].
\ee Finally the luminosity distance $\tilde{d}_L$ is (see \citep{Ellis, Kantowski2})
\be\label{dl-rw}
\tilde{d}_L=(1+z_n)(1+z)e^{\sigma}d_A=e^{\tau/2}\bar{d}_L.
\ee 
We have thus obtained the same luminosity-redshift relation we obtained in \citep{Kantowski2}  
for the real eikonal case (refer to Eq.\,(66) of \citep{Kantowski2}) by changing the geometry 
rather than absorbing some of the wave's intensity.
Just as in \citep{Kantowski2}, Eq. (\ref{dl-rw})  differs from the standard reciprocity 
relation \citep{Etherington, Bertotti, Ellis}
\be
d_L(z)=(1+z)^2 d_A(z),
\ee 
in two aspects: first, the existence of refraction causes photon orbits to 
deviate from null geodesics in the physical spacetime, this giving the $1+z_n$ 
factor instead of $1+z$; second, light absorption (expressed by the conformal 
factor $\sigma$) violates the photon number conservation law. For an interesting discussion about using  
reciprocity relation as a probe of acceleration/exotic physics, see \citep{Bassett}. 
\section{An Application: FLRW Spacetime with Both Refraction and Absorption \label{sec:numerical}}

In \cite{Kantowski1,Kantowski2} we have interpreted the observed apparent increase 
in the universe's expansion rate as caused by light refraction and absorption respectively, 
instead of by a cosmological constant $\Lambda.$ In \cite{Kantowski1} we used a two parameter 
pure refraction model, \ie  $n(z)=1+pz^2+qz^3,$ without absorption to fit the supernovae gold 
sample \cite{Http}. In \cite{Kantowski2} 
we fit the sample with a pure absorption model, \ie we took $n=1$, 
and the opacity $\alpha(z)\equiv\rho(z)\kappa(z)={\rm const}.$ In this section, 
we fit the same data set \citep{Riess3, Riess4, Astier, Davis, Wood} 
with a cold dark matter model containing both refraction and absorption. 
We use simple expressions $n(z)= 1+ pz^2$ and $\alpha(z)={\rm const}$ 
that depend on only two parameters. Since we are concerned with the matter 
dominated era, we have excluded radiation ($\Omega_r=0)$ and since we are 
trying to only emulate acceleration we take $\Lambda=0.$ We choose the 
current Hubble constant to be $H_0=65\,{\rm km}/{\rm s}/{\rm Mpc}$ and 
fit our two parameter ($\alpha$,\, $p$) model for different choices of 
$\Omega_{\rm m}.$ The scaled Hubble function $h(z)$ in this case simplifies to
\be
h(z)=(1+z)\sqrt{1+\Omega_{\rm m} z}.
\ee 
 The refraction and absorption corrected distance-redshift relation is now written as
\be
\tilde{d}_L(z_n)=(1+z_n)\frac{c}{H_0}\frac{e^{\sigma(z_n)}}{\sqrt{|\Omega_k|}}{\rm sinn}\sqrt{|\Omega_k|}\int_{0}^{z(z_n)}
{\frac{dz'}{n(z')h(z')}},
\label{d-z}
\ee where 
\be
\sigma(z_n)={\alpha\over 2}{c\over H_0}\int_{0}^{z(z_n)}
{{dz'\over (1+z')n(z')h(z')}},
\ee and
\be
1+z_n=\frac{1+z}{1+pz^2}.
\ee

\begin{figure*}
\includegraphics{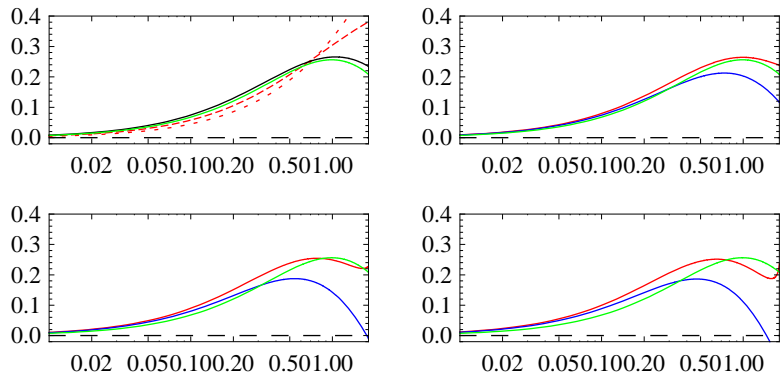}
\caption{$\Delta \mu$ versus $z.$ In each of the four frames the fiducial model 
(horizontal black dashed) is the now disfavored dark matter only model, 
\ie $\Omega_\Lambda=0,\Omega_{\rm m}=0.3,\alpha=0, p=0,$ and the green curve (gray in black \& white)
is the concordance model, $\Omega_\Lambda=0.7,\Omega_{\rm m}=0.3,\alpha=0, p=0.$ 
The upper left panel contains  pure absorption models with no refraction: 
the dotted red curve has $\Omega_\Lambda=0,\Omega_{\rm m}=0.05,\alpha=0.7\times 10^{-4}{\rm Mpc}^{-1}, n=1;$ 
the short-dashed red curve has $\Omega_\Lambda=0,\Omega_{\rm m}=0.3,\alpha=1.3\times 10^{-4}{\rm Mpc}^{-1}, n=1;$ 
the solid black curve has $\Omega_\Lambda=0,\Omega_{\rm m}=0.73,\alpha=2.2\times 10^{-4}{\rm Mpc}^{-1}, n=1.$ 
In the upper right panel: the blue (bottom) curve has $\Omega_\Lambda=0,\Omega_{\rm m}=1.0,\alpha=2.5\times 10^{-4}{\rm Mpc}^{-1}, p=0;$ 
the red (top) curve has $\Omega_\Lambda=0,\Omega_{\rm m}=1.0,\alpha=2.5\times 10^{-4}{\rm Mpc}^{-1}, p=0.024.$ 
In the bottom left panel: blue (bottom) curve has $\Omega_\Lambda=0,\Omega_{\rm m}=1.5,\alpha=3.3\times 10^{-4}{\rm Mpc}^{-1}, p=0;$ 
The red (top) curve has $\Omega_\Lambda=0,\Omega_{\rm m}=1.0,\alpha=3.3\times 10^{-4}{\rm Mpc}^{-1}, p=0.042.$ 
In the bottom right panel: blue (bottom) curve has $\Omega_\Lambda=0,\Omega_{\rm m}=2.0,\alpha=4.1\times 10^{-4}{\rm Mpc}^{-1}, p=0;$ 
the red curve has $\Omega_\Lambda=0,\Omega_{\rm m}=2.0,\alpha=4.1\times 10^{-4}{\rm Mpc}^{-1}, p=0.049.$  }  
\label{fig:Mu-Z}
\end{figure*}

\begin{figure*}
\includegraphics{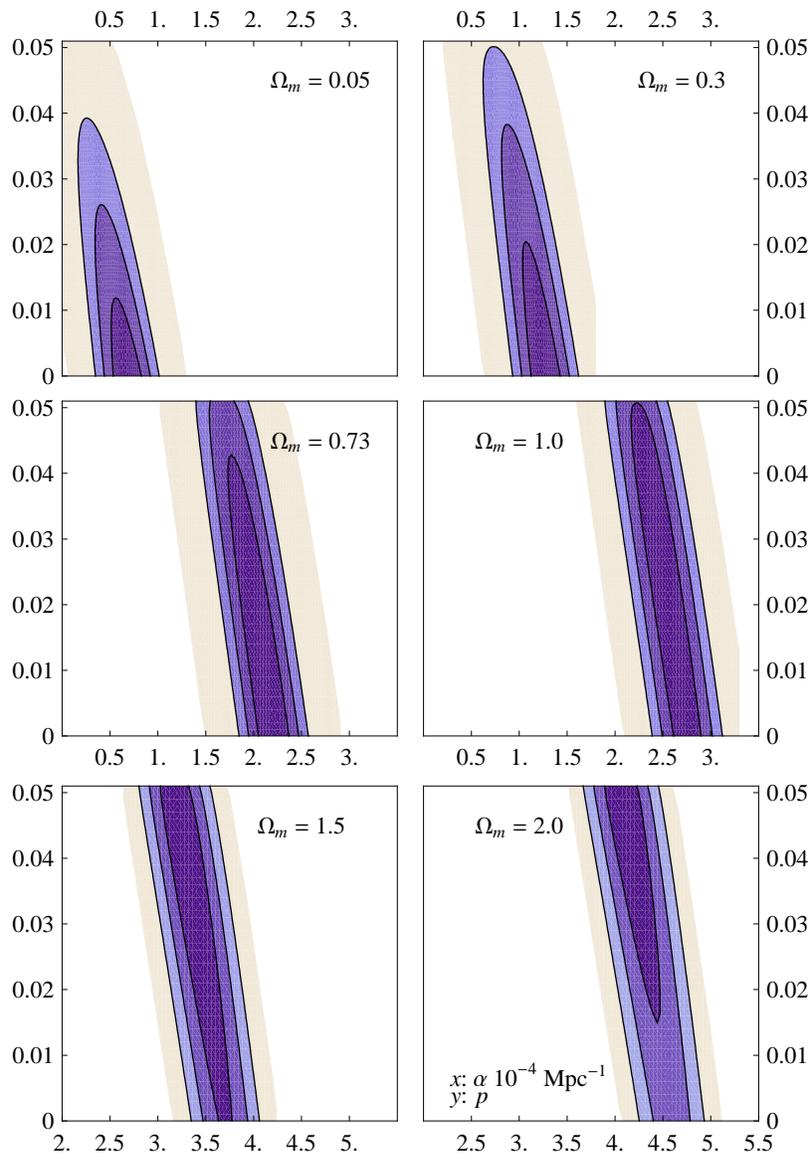}
\caption{Confidence contours ($68.3\%$, $95.4\%$, and $99.73\%$) for each fixed $\Omega_{\rm m}$ model. 
The $x$ coordinate is the absorption coefficient $\alpha$ in unit of $10^{-4}{\rm Mpc}^{-1},$ and 
the $y$-axis is the refraction index parameter $p$ ($n=1+pz^2$).  In the top left panel, 
$\Omega_{\rm m}=0.05,$ with least chi-square (per degree of freedom) $\chi^2=1.09,$ 
the best fitting parameters are $\alpha=0.7\times 10^{-4}{\rm Mpc}^{-1},\, p=0.$ 
In the top right panel, $\Omega_{\rm m}=0.3,$ least $\chi^2=1.05,$ the best fitting parameters 
are $\alpha=1.3\times 10^{-4}{\rm Mpc}^{-1},\, p=0.$ In the middle left panel, 
$\Omega_{\rm m}=0.73,$ least $\chi^2=1.035,$ the best fitting parameters 
are $\alpha=2.2\times 10^{-4}{\rm Mpc}^{-1},\, p=0.$ In the middle right panel, 
$\Omega_{\rm m}=1.0,$ least $\chi^2=1.042,$ at $\alpha=2.5\times 10^{-4}{\rm Mpc}^{-1},\, p=0.024.$ In the bottom left panel, 
$\Omega_{\rm m}=1.5,$ least $\chi^2=1.05,$ the best fitting parameters are 
$\alpha=3.3\times 10^{-4}{\rm Mpc}^{-1},\, p=0.042.$ I the bottom right panel, 
$\Omega_{\rm m}=2.0,$ least $\chi^2=1.06,$ the best fitting parameters 
are $\alpha=4.1\times 10^{-4}{\rm Mpc}^{-1},\, p=0.049.$  }
\label{fig:chi2}
\end{figure*}

We use the $178$ supernovae from the gold sample 
\cite{Http}
with redshifts greater than $cz = 7000\> \hbox{km}/\hbox{s}$ 
to avoid any Hubble bubble. Our results are shown in Figures \ref{fig:Mu-Z} and
\ref{fig:chi2}. In Fig.\,\ref{fig:Mu-Z}, we show the $\Delta\mu(z)$ curves for different model parameters. 
Here $\Delta\mu(z)\equiv \mu(z)-\mu_{F}(z),$ where $\mu_F(z)$ is the distance modulus of the fiducial, 
dark matter only model (horizontal black dashed curve in Fig.\,\ref{fig:Mu-Z}), 
\ie $\Omega_\Lambda=0,\Omega_{\rm m}=0.3,\alpha=0, p=0.$ In each of the four frames 
the green curve (grey in black \& white) is the concordance model, $\Omega_\Lambda=0.7,\Omega_{\rm m}=0.3,\alpha=0, p=0.$ 
In the upper left panel we show absorption models with no refraction index ($\Omega_{\rm m}<1$ 
for all curves in this panel). 
The dotted red curve has: $\Omega_\Lambda=0,\Omega_{\rm m}=0.05,\alpha=0.7\times 10^{-4}{\rm Mpc}^{-1}, n=1.$ 
The short-dashed red curve has: $\Omega_\Lambda=0,\Omega_{\rm m}=0.3,\alpha=1.3\times 10^{-4}{\rm Mpc}^{-1}, n=1.$ 
The solid black curve has:  $\Omega_\Lambda=0,\Omega_{\rm m}=0.73,\alpha=2.2\times 10^{-4}{\rm Mpc}^{-1}, n=1.$ In the upper right panel the models are flat. 
The blue curve (bottom curve) has: $\Omega_\Lambda=0,\Omega_{\rm m}=1.0,\alpha=2.5\times 10^{-4}{\rm Mpc}^{-1}, p=0.$ 
The red curve (top curve) has: $\Omega_\Lambda=0,\Omega_{\rm m}=1.0,\alpha=2.5\times 10^{-4}{\rm Mpc}^{-1}, p=0.024.$ 
In the bottom left panel  the blue curve (bottom curve) has: $\Omega_\Lambda=0,\Omega_{\rm m}=1.5,\alpha=3.3\times 10^{-4}{\rm Mpc}^{-1}, p=0.$ 
The red curve (top curve) has: $\Omega_\Lambda=0,\Omega_{\rm m}=1.5,\alpha=3.3\times 10^{-4}{\rm Mpc}^{-1}, p=0.042.$ 
In the bottom right panel the blue curve (bottom curve) has: $\Omega_\Lambda=0,\Omega_{\rm m}=2.0,\alpha=4.1\times 10^{-4}{\rm Mpc}^{-1}, p=0.$ 
The red (top curve) curve has: $\Omega_\Lambda=0,\Omega_{\rm m}=2.0,\alpha=4.1\times 10^{-4}{\rm Mpc}^{-1}, p=0.049.$ 

In Fig.\,\ref{fig:chi2} we show the confidence contours ($68.3\%$, $95.4\%$, and $99.73\%$) 
of our two parameter $(\alpha,p)$ model with different choices of $\Omega_{\rm m}.$ We show $6$ 
different cases with $\Omega_{\rm m}=0.05$ (baryonic matter only), $\Omega_{\rm m}=0.3$ 
(dark matter only), $\Omega_{\rm m}=0.73$ (our best least $\chi^2$ pure absorption model, 
see \cite{Kantowski2}), $\Omega_{\rm m}=1.0$ (flat universe), $\Omega_{\rm m}=1.5$ 
and $2.0$ (closed models). We restrict the parameter $p$ to be nonnegative to 
keep the light speed $v=c/n$ less than c. For the first three cases, 
\ie $\Omega_{\rm m}=0.05,0.3$ and $0.73,$ the best fits occur for $p=0,$ 
which suggests that introducing additional refraction parameters would not 
significantly improve the fitting when absorption is present. However, 
as $\Omega_{\rm m}$ increases, nonvanishing $p$ values do give better fits. 
For $\Omega_{\rm m}=1.0,$ the best fitting parameters ($\chi^2=1.04$) 
are $\alpha=2.5\times 10^{-4}{\rm Mpc}^{-1}, p=0.024.$ For $\Omega_{\rm m}=1.5,$ 
the best fitting ($\chi^2=1.05$) parameters are $\alpha=3.3\times 10^{-4}{\rm Mpc}^{-1}, 
p=0.042.$ For $\Omega_{\rm m}=2.0,$ the best fitting ($\chi^2=1.06$) 
parameters are $\alpha=4.1\times 10^{-4}{\rm Mpc}^{-1}, p=0.049.$ 
This can also be seen from Fig.\,\ref{fig:Mu-Z}: In the upper right 
and two bottom frames, the blue and red curves have the same absorption 
coefficient $\alpha,$ the difference is that the red curve has $p$ nonzero  
whereas the blue curve has $p=0.$ The inclusion of $p$ for these large 
$\Omega_{\rm m}$ cases improves the fits.    

The Supernova data is currently considered to be the most compelling evidence 
for the existence of dark energy because of the apparent 
acceleration observed in the expansion of the universe 
(see e.g.  \cite{Riess,Perlmutter}). 
Competing interpretations have been proposed, e.g., 
evolutionary effects \citep{Drell, Combes}, local Hubble bubbles \cite{Conley,Zehavi}, 
absorption \citep{Aguirre,Aguirre2,Rowan}, modified gravity \cite{Ishak,Kunz,Bertschinger} 
and others such as  slowly changing fundamental constants \cite{Albrecht,Barrow}.

For supernovae there are at least four different sources of opacity; the Milky Way, the hosting galaxy, 
intervening galaxies, and the IGM that should be taken into account. The Galactic 
absorption has been studied extensively \citep{Spitzer, Mathis} and
early constraints on properties of IGM were often obtained assuming 
the applicability of Galactic dust properties (see e.g. \citep{Wright}).  
The luminosity of high redshift supernovae have been corrected for host galaxy absorption 
using the Galactic reddening law, see \eg \cite{Riess, Riess1, Riess2, Perlmutter}. 
As has been pointed out caution should be exercised when applying Galactic dust properties to the IGM,  
since the composition, size, shape, and alignment of intergalactic 
dust could be significantly different than that of Milky Way dust. 
Aguirre \citep{Aguirre, Aguirre2} introduced a carbon needle model and 
showed that dust grains of larger size ($\geq 1\>\mu m$), which should be preferentially 
ejected by star burst galaxies, would have relatively higher opacities
(e.g. $\kappa\sim 10^5\> \hbox{cm}^2\, g^{-1}$) and much greyer 
absorption curves, and therefore might escape the reddening censorship 
based on Galactic reddening law. Evidence against this model  appeared in \citep{CroftA, Knop, Sullivan}. 
\citet{Goobar} introduced a replenishing dust model in which the dilution caused by cosmic expansion 
is continually replenished ($\rho=\rho_0(1+z)^3$ for $z<0.5,$ $\rho=\rho_0$ for $z\geq 0.5$). 
This dust model is indistinguishable from $\Omega_\Lambda$ and cannot be ruled out by 
Supernovae data alone (see table 5 of \citep{Riess4}). \citet{Bassett} claimed to rule 
out the replenishing model at more than $4\sigma$ by considering violations of distance 
duality (reciprocity relation). \citet{Ostman} further constrained the magnitude of grey dust 
absorption from Quasar colors and spectra and claimed that for a wide range of intergalactic 
dust models, extinction larger than $0.2\>\hbox{mag}$ is ruled out (see also \citep{Inoue,Corasaniti}). 
A new grey dust model is proposed in \citep{Robaina}.  More complicated and fine tuned dust 
models will keep emerging in the future until dark matter/energy has been identified, if in fact it exists.

\eject

\section{Discussion}\label{sec:discussion}

In this paper, we introduced the concept of a ``transverse" conformal transformation, 
which allowed us to equate the intensity reduction of light caused by absorption with a geometrical 
reduction caused by the expansion of the cross-sectional area of light ray bundles without 
changing other desirable properties of waves, such as  
frequency and wavelength. This application of conformal transformations is new.  
We used it to generalize Gordon's optical metric to include light absorption via such a 
conformal transformation. This generalization is fundamentally different from 
\cite{Kantowski2} in which we included light absorption by making Gordon's metric complex. 
In the complex metric formalism, we  distinguished two different cases: strong and 
weak absorption. The strong case is where a non-negligible amount of absorption occurs on a wavelength scale and the weak case is where absorption is significant only over multitudes of wavelengths. In the strong case the eikonal remains complex but in the weak case it can be taken as real. In this paper we were  able to replace the effects of weak absorption 
by a special conformal transformation. The disadvantage of this formalism 
is that the optical metric depends on the eikonal of the wave 
itself whereas in \citep{Kantowski2} it did not. We used this new optical metric to derive the absorption and 
refraction corrected distance-redshift relation for FLRW spacetimes  and 
obtained the same expression as in \citep{Kantowski2} for weak absorption. 
In \citep{Kantowski1} we fit supernovae data with a pure refraction model, and 
in \citep{Kantowski2} we fit it with a pure absorption model. 
In Sec.\,\ref{sec:numerical} we fit this data with a cosmological 
model possessing both refraction and absorption. We have shown that a single parameter 
polynomial approximation for the refraction index, \ie $n(z)=1+pz^2,$ together with a constant opacity 
parameter $\alpha$ fits the data well. More realistic $z$ dependent models for 
$\alpha$ and $n$ would be appropriate. We have not proposed a physical source for the needed 
refraction and/or absorption. What we have done here is to construct a geometry modification whose effects are 
equivalent to absorption. It is not a new absorption theory only a new way of looking at its effects.
It equates the decrease in intensity of a wave caused by absorption with a decrease caused by a 
change in the underlying spacetime, \ie a transverse conformal transformation. We used this 
modified geometry to compute the distance-redshift relations in FLRW and applied it to the 
Hubble curve for type SNe-Ia.  In this application we ignored effects of inhomogeneities including the
possibility  of having a multi-valued conformal factor. We assumed that none of the SN Ia used 
were strongly lensed. 
Since the flat concordance model is consistent with other observations, e.g., the angular position of the first acoustic peak as measured by WMAP (Wilkinson Microwave Anisotropy Probe) team \citep{Spergel, Spergel2}, and baryonic acoustic oscillations (BAO) detected in galaxy surveys \citep{Tegmark, Tegmark2, Eisenstein}, 
an absorption and/or refraction theory cannot be on firm ground unless it 
is also consistent with these additional observations. We leave these 
and other applications to future efforts. 
    
\section{Acknowledgments}

This work was supported in part by NSF grant AST-0707704 and US DOE
Grant DE-FG02-07ER41517.

\label{lastpage}


\begin{thebibliography}{breitestes Label}

\bibitem[Gordon(1923)]{Gordon} W. Gordon, Ann. Phys. (Leipzig) 72, 421 (1923).

\bibitem[Chen \& Kantowski (2009)]{Kantowski2} B. Chen and R. Kantowski, Phys. Rev. D 79, 104007 (2009).

\bibitem[Eisenhart (1926)]{Eisenhart} L. P. Eisenhart, {\it Riemannian Geometry} (Princeton Univ., Princeton, 1926).

\bibitem[Penrose(1967)]{Penrose} R. Penrose, {\it An Analysis of the Structure of Space-Time, Adams Prize Essay} (Princeton Univ., Princeton, 1967).

\bibitem[Ehlers(1967)]{Ehlers2} J. Ehlers, Z. Naturforsch. 22a, 1328 (1967).   
  
\bibitem[Sachs(1961)]{Sachs0} R. Sachs, Proc. R. Soc. A 264, 309 (1961).

\bibitem[Mihalas(1970)]{Mihalas} D. Mihalas, {\it Stellar Atmospheres} (W. H. Freeman and Company, San Fransisco, 1970).

\bibitem[Ellis(1971)]{Ellis} G. F. R. Ellis, in {\it General Relativity and Cosmology}, edited by R. K. Sachs (Academic, New York, 1971), p 104.

\bibitem[Ehlers(1966)]{Ehlers1} J. Ehlers, in {\it Perspectives in Geometry and Relativity}, edited by B. Hoffmann (Indiana Univ., Indiana, 1966), p. 127.



\bibitem[Etherington(1933)]{Etherington} I. M. H. Etherington, Philos. Mag. 15, 761 (1933).

\bibitem[Bertotti (1966)]{Bertotti} B. Bertotti, Proc. R. Soc. A. 294, 195 (1966).

\bibitem[Bassett \& Kunz (2004)]{Bassett} B. A. Bassett \& M. Kunz, Phys. Rev. D 69, 101305(R) (2004).

\bibitem[Chen \& Kantowski (2008)]{Kantowski1} B. Chen and R. Kantowski, Phys. Rev. D 78, 044040 (2008).

\bibitem[Riess \etal(2004)]{Riess3} A. G. Riess et al., Astrophys. J. 607, 665 (2004).

\bibitem[Riess \etal(2007)]{Riess4} A. G. Riess et al., Astrophys. J. 659, 98 (2007).

\bibitem[Astier \etal(2006)]{Astier} P. Astier et al., Astron. Astrophys. 447, 31 (2006). 

\bibitem[Davis \etal(2007)]{Davis} T. M. Davis et al., Astrophys. J. 666, 716 (2007).

\bibitem[Wood-Vasey \etal(2007)]{Wood} W. M. Wood-Vasey et al., Astrophys. J. 666, 694 (2007).

\bibitem[Http \etal(2006)]{Http} http://braeburn.pha.jhu.edu/$\sim$ariess/R06/. 

\bibitem[Riess \etal (1998)]{Riess} A. G. Riess et al., Astron. J. 116, 1009 (1998).

\bibitem[Perlmutter \etal (1999)]{Perlmutter} S. Perlmutter et al., Astrophys. J. 517, 565 (1999).

\bibitem[Drell \etal (2000)]{Drell} P. S. Drell, T. J. Loredo, and I. Wasserman, Astrophys. J. 530, 593 (2000).

\bibitem[Combes (2004)]{Combes} F. Combes,  New Astron. Rev.  48, 583 (2004).

\bibitem[Conley \etal (2007)]{Conley} A. Conley et al., Astrophys. J. 664, L13 (2007).

\bibitem[Zehavi \etal (1998)]{Zehavi} I. Zehavi et al., Astrophys. J. 503, 483 (1998).

\bibitem[Aguirre (1999)]{Aguirre} A. N. Aguirre, Astrophys. J. 512, L19 (1999).

\bibitem[Aguirre (1999)]{Aguirre2} A. N. Aguirre, Astrophys. J. 525, 583 (1999).

\bibitem[Rowan-Robinson (2002)]{Rowan} M. Rowan-Robinson, Mon. Not. R. Astron. Soc. 332, 352 (2002).

\bibitem[Ishak \etal (2006)]{Ishak} M. Ishak, A. Upadhye \& D. N. Spergel, Phys. Rev. D 74, 043513 (2006). 

\bibitem[Kunz \& Sapone (2007)]{Kunz} M. Kunz \& D. Sapone, Phys. Rev. Lett 98, 121301 (2007).

\bibitem[Bertschinger \& Zukin (2008)]{Bertschinger} E. Bertschinger \&  P. Zukin, Phys. Rev. D 78, 024015 (2008).

\bibitem[Albrecht \& Magueijo (1999)]{Albrecht} A. Albrecht and J. Magueijo, Phys. Rev. D 59, 043516 (1999).

\bibitem[Barrow (1999)]{Barrow} J. D. Barrow, Phys. Rev. D 59, 043515 (1999).

\bibitem[Spitzer(1978)]{Spitzer} L. Spitzer, {\it Physical Processes in the Interstellar Medium} (John Wiley \& Sons,  New York, 1978).

\bibitem[Mathis (1990)]{Mathis} J. S. Mathis, Ann. Rev. Astron. Astrophys. 28, 37 (1990).

\bibitem[Wright (1981)]{Wright} E. L. Wright,  Astrophys. J. 250, 1 (1981).

\bibitem[Riess \etal (1996)]{Riess1} A. G. Riess, W. H. Press, and R. P. Kirshner, Astrophys. J. 473, 88 (1996).

\bibitem[Riess \etal (1996)]{Riess2} A. G. Riess, W. H. Press, and R. P. Kirshner, Astrophys. J. 473, 588 (1996).

\bibitem[Croft \etal (2000)]{CroftA} R. A. C. Croft,  R. Dav$\acute{\rm e}$, L. Hernquist and N. Katz, Astrophys. J. 534, L123 (2000).

\bibitem[Knop \etal (2003)]{Knop} R. Knop et al.   Astrophys. J. 598, 102 (2003).

\bibitem[Sullivan \etal (2003)]{Sullivan} M. Sullivan et al.  Mon. Not. R. Astron. Soc. 340, 1057 (2003).

\bibitem[Goobar \etal (2002)]{Goobar} A. Goobar, L. Bergstr$\ddot{\rm o}$m, and E. M$\ddot{\rm o}$rtsell, Astron. Astrophys. 384, 1 (2002).


\bibitem[$\ddot{\rm O}$stman \& M$\ddot{\rm o}$rtsell  (2005)]{Ostman} L. $\ddot{\rm O}$stman \& E. M$\ddot{\rm o}$rtsell, J. Cosmol. Astropart. Phys. 2, 005 (2005).

\bibitem[Inoue \& Kamaya (2004)]{Inoue} A. K. Inoue \& H. Kamaya,  Mon. Not. R. Astron. Soc. 350, 729 (2004).

\bibitem[Corasaniti (2006)]{Corasaniti} P. S. Corasaniti,  Mon. Not. R. Astron. Soc. 372, 191 (2006).

\bibitem[Robaina \& Cepa (2007)]{Robaina} A. R. Robaina \& J. Cepa, Astron. Astrophys. 464, 465 (2007). 

\bibitem[Spergel \etal(2003)]{Spergel} D. N. Spergel et al. Astrophys. J. Suppl. Ser. 148, 175 (2003). 

\bibitem[Spergel \etal(2007)]{Spergel2} D. N. Spergel et al. Astrophys. J. Suppl. Ser. 170, 377 (2007). 

\bibitem[Tegmark \etal (2004)]{Tegmark} M. Tegmark et al. Phys. Rev. D 69, 103501 (2004). 

\bibitem[Tegmark \etal (2004)]{Tegmark2} M. Tegmark et al. Astrophys. J. 606, 702 (2004).

\bibitem[Eisenstein \etal (2005)]{Eisenstein} D. J. Eisenstein et al. Astrophys. J. 633, 560 (2005).


\end{thebibliography}
\end{document}